# Analiza bezbednosnih mehanizama OSPF protokola

Petar Bojović, Katarina Savić

*Sadržaj* — Bezbednost nekog servisa i sistema zavisi od bezbednosti svake komponente tog sistema. Napad na protokol rutiranja može proizvesti neispravno funkcionisanje računarske mreže. U pojedinim slučajevima moguće je da napadač dođe do podataka ili umetne podatke za koje nema pravo. OSPF protokol je najrasprostranjeniji protokol stanja linka (link-state protocol). U ovom radu izvršili smo analizu bezbednosti OSPF protokola i opisali njegove bezbednosne mehanizme.

*Ključne reči* — Bezbednost, link-state, MD5, Mikrotik, OSPF, XORP

## I. Uvod

Potrebe današnjih servisa elektronskog poslovanja diktiraju zahteve za visoku pouzdanost i bezbednost celokupnog sistema. Računarske mreže su ključan resurs u ispunjenju funkcionalnosti tih servisa. Bezbednost računarskih mreža se analizira kroz dobro poznat model računarskih mreža – TCP/IP model. Prava zaštita se dobija tek implementacijom zaštitnih mehanizama na svim nivoima mrežnog modela [1].

Na fizičkom sloju zaštita se dobija pažljivom fizičkom kontrolom prustupa mreži, skrivanjem pristupnih tačaka, i mrežnih portova. Kod bežičnih mreža pažljivim izborom tipova antena, tj. izborom usmerenih na uštrb omni-direkcionih [1], [2].

Na sloju veze zaštita se obezbeđuje filtriranjem MAC adrese ili drugih identifikatora. Važnu ulogu ima izbor autentifikacionih metoda koje sprečavaju prijavljivanje neovlašćenih lica, upotreba VLAN-ova, L2 VPN. Kod bežičnih mreža bitna je autentifikacija preko deljenog ključa, MAC adrese, ili PKI sistema, kao i enkripcija WEP, WPA [2].

Na sloju mreže zaštita se postiže filtriranjem IP adrese, upotrebom L3 VPN

Peter Bojović, Računarski fakultet univerziteta Union, Knez Mihailova 6/VI, 11000 Beograd, Srbija (telefon: +381-11-2633-321; e-mail: raf@paxy.in.rs).
Katarina Savić, Elektrotehnički fakultet, Univerzitet u Beogradu, (e-mail: catrins@gmail.com).

tunela, i zaštitom ruting protokola od neovlašćene izmene [2].

Na transportnom sloju mehanizmi zaštite podrazumevaju implementaciju zaštitnih barijera (firewall) koji sprečavaju pakete koji se šalju na portove za koje nije predviđena usluga [1].

Na aplikativnom sloju postoje razni metodi zaštite tipa, SSL VPN, upotreba simetričnih i asimetričnih kriptografskih algoritama, PKI, heš algoritama i slično [5], [4].

Različitim mrežnim servisima je potreban različiti nivo bezbednosti. Povećavanjem nivoa bezbednosti nekog servisa smanjuje se udobnost i fleksibilnost pri radu sa tim servisom. Na primer, servis koji je potpuno zaštićen je teško koristiti, jer se od korisnika zahteva da izvrši čitav niz operacija pre korišćenja servisa, pa da se potom striktno pridržava pravila korišćenja. Ovakvi servisi uglavnom brzo budu odbačeni od strane korisnika, tj. neophodno je da servis pruža dosta koristi kako bi korisnik prihvatio proceduru njegovog korišćenja. Zbog toga je uvek neophodno napraviti kompromis između nivoa bezbednosti i lakoće pri korišćenju servisa. Taj kompromis se ogleda u implementaciji samo pojedinih, transparentnih, sistema zaštite.

Na mrežnom sloju TCP/IP modela se mogu implementirati određeni nivoi zaštite koji nikako ne opterećuju korisnika servisa, već se postavljaju prilikom projektovanja mreže. Prilikom postavljanja računarske mreže, dizajneri te mreže postavljaju i prava korišćenja protokola rutiranja u toj mreži. Jedan od najrasprostranjeniji protokola rutiranja je OSPF (Open Shortest Path First).

## II. RANJIVOST OSPF PROTOKOLA

Kao otvoreni standard, OSPF protokol poseduje mnogo kritika na račun određenih implementacija i metoda. Neki od napada na OSPF protokol, kao i moguća rešenja protiv tih napada su objavljeni u [3]. Pre svega trebalo definisati nekoliko pojmova. Napadač je onaj/ono koji ne ovlašćeno vrši izmene ili na bilo koji drugi način utiče na rad OSPF protokola [3].

Jedna vrsta napada na protokole rutiranja potiče od namere da se preusmeri tok kretanja paketa preko rutera koji nisu planirani dizajnom mreže. Na taj način se preko tih neovlašćenih rutera može vršiti posmatranje i snimanje svog ili drugog interesantnog saobraćaja na mreži. Napadač bi preusmeravanjem mogao da omogući da se na neovlašćenom ruteru vrši beleženje svih paketa, pa analizom tih paketa do dođe u posed podataka koji se prenose mrežom.

Drugi razlog napada na protokol rutiranja može biti da se privremeno, ili potpuno onemogući protok paketa kroz mrežu. Time bi se srušio protokol rutiranja i onemogućila komunikacija na mrežnom sloju. Treći razlog napda može da bude slučajno onesposobljavanje mreže koje se može desiti kada se vrši ovlašćena intervencija na mreži, ali se nisu sagledale sve posledice te intervencije.

Tehnike napada se mogu svrstati u sledeće kategorije. Prisluškivanje koje se odnosi na to da ako se podaci o rutama prenose u čitljivoj ne kodiranoj formi, onda je moguće da napadač jednostavnim prisluškivanjem i analizom dođe do informacija kako izgleda topologija mreže koju napada. Odgovor na poruku se odnosi na mogućnost napadača da kada detektuje poruku koja mu može omogućiti izvršenje željenog napada, izvrši odgovor na poruku drugog rutera i time se ubaci u OSPF proces. Ubacivanje poruke je metod ubacivanja poruke u mrežni tok sa ciljem izazivanja željenog napada. Brisanje poruke se odnosi na postupak kojim se ometanjem poslate poruke može pokušati izazvati prekid relacije između dva učesnika OSPF procesa. Modifikacija poruke je izmena postojeće poruke sa ciljem lažnog predstavljanja radi izvršenja napada. Čovek u sredini (man-in-the-middle) je napad prilikom kog se napadač postavlja između dva učesnika OSPF procesa, transparentno učestvujući neovlašćeno u OSPF procesu. Odbijanje servisa (denial-of-service) je napad koji ima za cilj da ubaci veliki broj beskorisnih paketa o rutiranju tako da spreči ispravno tumačenje pravih paketa o rutiranju. Cilj ovog napada je onesposobljavanje mreže.

### III. ZAŠTITA OSPF PROTOKOLA

OSPF protokol je zbog svoje otvorene implementacije danas deo i najjednostavnijih rutera. To je protokol koji ima široku primenu. Pomenuti napadi i tehnike napada se mogu preventivno sprečiti korišćenjem sigurnosnih mehanizama koje su propisane OSPF protokolom i dodatnom zaštitom na mrežnom i nižim slojevima (sloj veze, fizički sloj). U okviru OSPF standarda postoje definisane dve metode zaštite OSPF procesa koje su objašnjene u nastavku.

*A. Jednostavna autentifikacija lozinkom* (*Simple password authentication*)

Metod proste autentifikacije lozinke predviđa prenos lozinke u tekstualnom obliku u okviru svake OSPF poruke, bilo da se radi o *Hello* ili *LSA* paketima, u posebnom okviru zaglavlja [4]. Dužina lozinke koja se može koristiti za ovaj način autentifikacije je do 64 bita to jest 8 karaktera.

Da bi se postigla autentifikacija na ovaj način, svi ruteri koji učestvuju u

OSPF procesu moraju imati identičnu lozinku u svojoj konfiguraciji. Prilikom slanja paketa lozinka se ubacuje u odgovarajuće polje svakog paketa. Prilikom prijema paketa proverava se polje sa lozinkom i poredi sa lozinkom u konfiguracionom fajlu. Paket se obrađuje samo ukoliko se lozinke poklapaju.

Metoda zaštite prenošenjem lozinke na ovaj način štiti OSPF proces od rutera koji žele neovlašćen pristup OSPF procesu. Ovi ruteri neće uticati na sam proces dokle god nemaju odgovarajuću lozinku. Veliki nedostatak ovog metoda autentifikacije je to što svako ko ima mogućnost da se prisluškuje fizički, tj. sloj veze, može analizom paketa doći do informacija koja se lozinka koristi u OSPF procesu, pa tako i lako lažirati ispravnu autentifikaciju napadačkog rutera.

### B. *Kriptografska autentifikacija*

Alternativni način autentifikacije jeste korišćenjem nekog od kriptografskih algoritama autentifikacije [4]. Standardom je propisano da su najpogodniji kriptografski algoritmi za autentifikaciju poznati heš algoritmi. Za potrebe kriptografske autentifikacije koristi se MD5 heš algoritam [4], [5].

Prilikom konfiguracije mehanizma kriptografske autentifikacije OSPF procesa postavlja se lozinka na sve uređaje koji treba da učestvuju u zajedničkom OSPF procesu. Za razliku od jednostavne autentifikacije, ta lozinka se ne prenosi svakom porukom, ali se ta lozinka koristi prilikom generisanja ili provere heš vrednosti. Heš funkcija generiše jedinstveni otisak paketa i lozinke. Iz tog otiska je nemoguće regenerisati paket ili lozinku. Ovako dobijen otisak se šalje u svakom paketu kao autentifikacioni parametar.

Algoritam MD5 se koristi kao heš algoritam. On proizvodi otisak dužine 16 bajtova bez obzira na veličinu paketa nad kojim se izvršava heš funkcija. Prilikom generisanja paketa, u polje za autentifikaciju se upisuje tzv, *null* lozinka. Zatim se taj paket zajedno za deljenom lozinkom ubacuje u MD5 algoritam i dobija se 16 bajtni otisak. Taj otisak se zatim ubacuje u polje za autentifikaciju i šalje. Prilikom prijema paketa, izvuče se parametar iz polja za autentifikaciju, u to polje upiše *null* vrednost, zajedno sa deljenom lozinkom se izračuna MD5 otisak, i proveri da li je izračunata i primljena vrednost identična.

Ukoliko se MD5 vrednosti poklapaju to znači dve stvari. Prvo, korišćen je isti deljen ključ, lozinka, i na jednoj i na drugoj strani. Drugo, paket nije promenjen između slanja i prijema. Ukoliko se vrednosti ne poklapaju takav paket se odbacuje i ne procesira.

Korišćenjem jedne statičke lozinke, deljenog ključa, povećava se verovatnoća da će neko pre ili kasnije doći do te lozinke. Kao rešenje uvodi se pojava dinamičkih ključeva, tj. mogućnost da se unese više ključeva na sve rutere pa da se u zavisnosti od vremena koristi jedan ili drugi. Tako na primer, dva sata važi jedan ključ, pa dva drugi, pa dva treći i tako dalje. Ovo eliminiše mogućnost da će neko saznati deljeni ključ, ali i donosi probleme vezane za distribuciju tih ključeva. Ovo se može rešiti tako da se prilikom konfiguracije ubacuju svi ključevi na sve uređaje jednog OSPF procesa sa vremenskom odrednicom kada se koriste. Dodatni problem predstavlja sinhronizacija vremena, jer svi ruteri moraju imati identično podešen sat kako bi se znalo koji ključ da koriste. Rešenje je korišćenje protokola za sinhronizaciju vremena, NTP protokola. Indeks ključa omogućava ruterima da se sporazumeju koji ključ treba koristiti.

Kriptografskom autentifikacijom je sprečen upad u OSPF proces lažnim predstavljanjem kao legitiman ruter zbog toga što se šifra nikad ne prenosi pa ni napadač ne može lako doći do te šifre. Kao dodatna mera bezbednosti dodat je i specifičan brojač rednih brojeva „*Cryptographic sequence number*" [4]. Inkrementacijom ovog brojača nakon svake razmene paketa, postiže se to da se sprečava napad tipa Odgovor na poruku, jer se paketi sa istim rednim brojem ne obrađuju, a ukoliko obe strane primete poremećaj u broju sekvence, dolazi do raskidanja i reiniciranja odnosa. Tehnika napada Ubacivanje poruke je sprečena tako što ukoliko se redni broj ne podudara sa očekivanom, poruka se ignoriše. Modifikacija poruke je sprečena korišćenjem MD5 algoritma, jer modifikovana poruka sigurno neće imati isti otisak kao i originalna. Napad čovek u sredini je onemogućen iz tog razloga što lozinka nikad ne napušta konfiguraciju rutera, a bez nje napadač ne može da učestvuje u OSPF procesu.

Napad Odbijanje servisa se ne može rešiti korišćenjem kriptografske autentifikacije. Ovaj problem zahteva korišćenje naprednih alata na svim nivoima mrežnog modela. Jedini efikasan način za prevenciju ovog napada jeste korišćenjem sistema za detekciju i prevenciju tj. IDS (Intrusion detection system) i IPS (Intrusion prevention system) koji vrše stalno nadgledanje mreže i kada detektuju napad, izvrše niz modifikacija na ruteru i zaštitnoj barijeri sa ciljem da spreče dalji napad. Detekcija se vrši na taj način što se broje paketi i konekcije koje se uspostavljaju po OSPF protokolu, i kada broj pređe određenu granicu to se smatra DOS napadom. Tada bi zaštitni zid trebalo da izvrši blokadu prenosa tih paketa, filtrirajući ih po MAC ili IP adresi.

Napad Prisluškivanja predstavlja tip napada namenjen za izviđanje. Njime se teži da onaj ko ne poznaje topologiju mreže, sazna kako ta mreža izgleda i time lakše „nađe žrtvu". Autentifikacioni metodi kao što su metod jednostavne autentifikacije lozinke ili MD5, ne pružaju tajnost prenetih paketa. Zato svako ko može da pristupi fizičkom i sloju veze može da pročita sastav paketa i analizom dobije izgled topologije mreže. Da bi se obezbedila tajnost paketa između uređaja u OSPF procesu mora se koristiti neki od metoda za tunelovanje. OSPFv3 je OSPF standard prilagođen za IPv6 koji po standardu predviđa korišćenje IPSec tunelovanje. IPSec se može koristiti i sa OSPFv2 ali njegovo korišćenje nije predviđeno standardnom. Ukoliko se koristi IPSec u OSPFv2 prvo se podiže IPSec tunel između susednih rutera, pa onda OSPF proces. Kako sam standard nije obuhvatio IPSec tako je neophodno prilikom podizanja tunela posebno podesiti parametre za Autentifikaciju i Enkripciju. IPSec protokol za tunelovanje podržava različite metode autentifikacije. Najkompleksnija je metoda upotrebom PKI sistema, koja omogućava korišćenje kripto modula u cilju vrhunske bezbednosti.

IPSec je tehnologija trećeg sloja (L3) za ostvarenje virtelnih privatnih mreža (VPN) koja podržava prenos paketa za rutiranje. Tunel može biti uspostavljen i drugim L3 VPN tehnologijama kao što su: PPTP, L2TP, i slično. Korišćenjem tunela svi paketi koji prolaze kroz tunel su šifrovani sesijskim ključevima koji se menjaju prilikom svake uspostave. Napadač koji prisluškivanjem pokušava da dobije neku korisnu informaciju dobiće niz paketa koji ništa ne znače bez sesijskog ključa.

### IV. IMPLEMENTACIJA BEZBEDNOSNIH MEHANIZAMA KOD RAZLIČITIH PROIZVOĐAČA OPREME

Sama implementacija autentifikacionih mehanizama se razlikuje kod različitih proizvođača opreme. Iako su implementacije različite, uglavnom oprema različitih proizvođača može da učestvuje u istom OSPF procesu.

#### A. CISCO implementacija OSPF bezbednosti

CISCO IOS operativni sistem koji se izvršava na ruteru predviđa prilikom konfiguracije OSPF procesa konfiguraciju dva moda: *router* i *interface*.

Unutar *router* konfiguracionog moda podiže se OSPF proces i postavljaju parametri koji su karakteristični za OSPF proces. Neophodno je u okviru istog moda oglasiti i mreže koje učestvuju u OSPF procesu, kao što je prikazano u primeru 1

```
Router>
Router>enable
```

```
Router#configure terminal
Router(config)#router ospf 1234
Router(config-router)#network 192.168.1.4 0.0.0.3 area 0
```
*Primer 1. – Oglašavanje mreža u OSPF proces*

U primeru 1, komandom *router ospf 1234* pokrećemo OSPF proces sa internim identifikatorom procesa 1234. Komandom *network 192.168.1.4 0.0.0.3 area 0* oglašavamo mrežu 192.168.1.4/30 u Backbone area tj. area 0.

Konfiguracija u *interface* modu se razlikuje u zavisnosti od toga da li želimo da postavimo jednostavnu autentifikaciju lozinkom ili kriptografsku autentifikaciju.

```
Router(config)#interface fa 0/0
Router(config-if)#ip address 192.168.1.5 255.255.255.252
Router(config-if)#ip ospf authentication
Router(config-if)#ip ospf authentication-key nekasifra
```
*Primer 2. – Postavljanje jednostavne autentifikacije lozinkom*

Nakon postavljanja IP adrese na odgovarajućem interfejsu, potrebno je podići autentifikaciju za OSPF proces. Komandom *ip ospf authentication* definišemo da će se koristiti jednostavna autentifikacija lozinkom. Komandom *ip ospf authentication-key nekasifra* postavljamo lozinku na „nekasifra", primer 2.

Ukoliko koristimo Kriptografsku autentifikaciju konfiguracija interfejsa se vrši na sledeći način:

```
Router(config)#interface fa 0/0
Router(config-if)#ip address 192.168.1.5 255.255.255.252
Router(config-if)#ip ospf authentication message-digest
Router(config-if)#ip ospf message-digest-key 16 md5 ovojesifra
```
*Primer 3. – Postavljanje MD5 kriptografske autentifikacije*

Nakon postavljanja IP adrese na odgovarajućem interfejsu, potrebno je podići autentifikaciju za OSPF proces. Komandom *ip ospf authentication message-digest* definišemo da će se koristiti MD5 kriptografska autentifikacija. Komandom *ip ospf message-digest-key 16 md5 ovojesifra* definišemo ključ sa rednim broj 16 tako da u MD5 algoritam sa paketom ulazi lozinka „ovojesifra", primer 3.

B. Mikrotik implementacija OSPF bezbednosti

Kod Mirotik rutera konfiguracija OSPF procesa se može izvršiti preko grafičkog interfejsa Winbox koji predstavlja Windows GUI za Mikrotik konfiguraciju. Po pokretanju aplikacije unosimo IP adresu i autentifikacione parametre za pristup MT uređaju.

U listi menija izborom *Routing* opcije može se konfigurisati OSPFv2 i OSPFv3 protokol. Unutar OSPF opcije otvara se okvir sa opcijama za OSPF protokol. Izborom opcije *New OSPF* kreiramo OSPF proces i postavljamo parametre procesa. Najveći broj opcija ponuđenih za konfiguraciju OSPF-a se nalaze u okviru za kreiranje novog procesa, pogledati Sliku 1. Tu se postavljaju parametri kao što su: na kom interfejsu da se izvršava, koja je podrazumevana cena rute, prioritet za DR, BDR izbor, tip autentifikacije, tip mreže, da li je proces aktivan na tom interfejsu, kao i tajmeri za validaciju paketa.

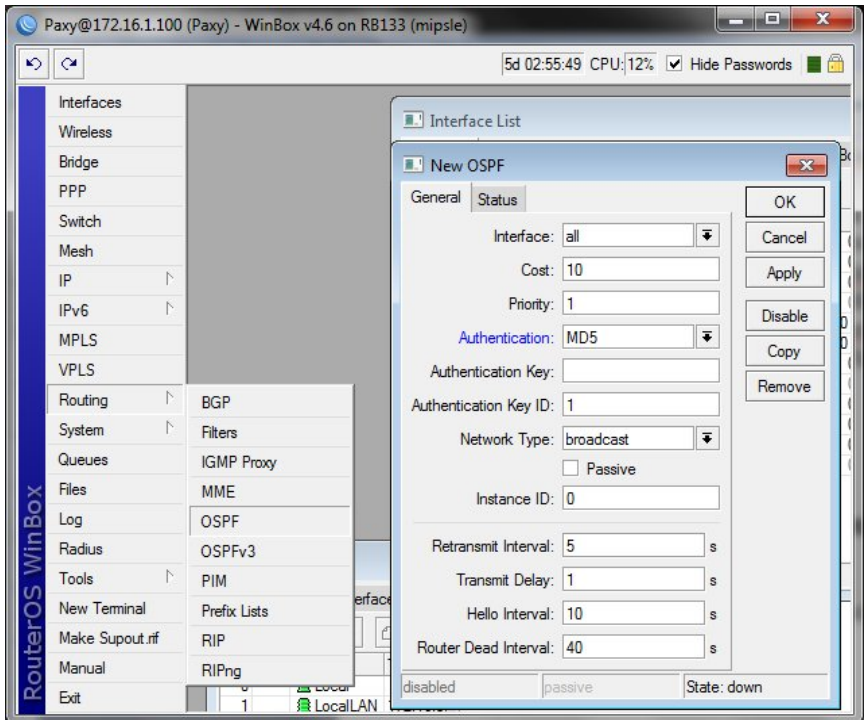

*Slika 1. – Mikrotik GUI okvir za OSPF autentifikaciju*

Autentifikacija može biti: None, kad se ne koristi autentifikacija, Simple kada se koristi slanje lozinke iz polja Authentication key, MD5 kada se vrši kriptografska autentifikacija koristeći lozinku Authentication key sa rednim brojem ključa iz polja Authentication key ID. Oglašavanje mreže se vrši u okviru Network na OSPF formi.

### C. XORP implementacija OSPF bezbednosti

XORP platforma poseduje proširenu podršku za Kriptografsku autentifikaciju dinamičkim ključevima. XORP podržava OSPFv2 i OSPFv3. Pre podizanja OSPF procesa potrebno je obezbediti IP adrese i omogućiti unicast rutiranje.

```
interfaces {
        restore-original-config-on-shutdown: false
        interface dc0 {
          description: "data interface"
          disable: false
          /* default-system-config */
          vif dc0 {
              disable: false
              address 10.10.10.10 {
                prefix-length: 24
                broadcast: 10.10.10.255
              disable: false
                }
              }
        }
    }
```
*Primer 4. – Podešavanje IP adrese na XORP platformi*

Podešavanje IP adrese u XORP platformi se vrši prema primeru 4. Ukoliko želite da koristite IP adresu koja je definisana u OS potrebno je skinuti komentare komande *default-system-config* i obrisati *vif* deklaraciju.

```
fea {
  unicast-forwarding4 {
    disable: false
    }
  }
plumbing {
  mfea4 {
    disable: false
    interface dc0 {
    vif dc0 {
      disable: false
    }
    }
    interface register vif {
      vif register vif {
      /* Note: this vif should be always enabled */
```

```
        disable: false
        }
      }
    traceoptions {
      flag all {
      disable: false
      }
    }
  }
```

*Primer 5. – Konfiguracija koja omogućuje IP unicast forwarding*

```
protocols {
    ospf4 {
        router-id: 10.10.10.10
          area 0.0.0.0 {
             interface dc0 {
                vif dc0 {
                   address 10.10.10.10 {
                      authentication {
                   /* simple-password: ovojeplainsifra */

                         md5 16 {
                              password: ovojemd5sifra
                              start-time: 2010-02-20:10:00
                              end-time: 2012-02-20:10:00
                              max-time-drift: 5
                              }
                              }
                      }
                   }
                }
             }
          }
       }
    }
```

*Primer 6. – Konfigurisanje OSPF procesa i autentifikacije na XORP platformi*

U primerima 5 i 6 je opisana konfiguracija OSPFv2 procesa, za ruterId 10.10.10.10 koji je član Backbone area, tj. area 0. Ta konfiguracija koristi MD5 autentifikaciju koristeći ključ ID 16, sa lozinkom „ovojemd5sifra" koji je validan od 20.2.2010 od 10h do 20.2.2012 u 10h. Tolerancija na desinhronizaciju u vremenu je 5 sekundi. Ukoliko se želi jednostavna autentifikacija lozinkom potrebno je skinuti komentar sa komande *simple-password* i obrisati *md5* sekciju.

ABSTRACT

Service or system security depends on the security of any component used on that system. Computer network attacks can jeopardize normal network functionality. There are cases where an attacker can gain unauthorized control over classified data. OSPF is the most common link state routing protocol. In this paper, we have analyzed OSPF security issues and described some protection methods.

**Analysis of OSPF security mechanisms**
Petar Bojović, Katarina Savić